\documentclass{aastex6}

\fullcollaborationName{The Friends of AASTeX Collaboration}

\begin{document}


\title{Data Compression for the Tomo-e Gozen with Low-rank Matrix Approximation}


\author{Mikio Morii\altaffilmark{1},
Shiro Ikeda\altaffilmark{1},
Shigeyuki Sako\altaffilmark{2}
and 
Ryou Ohsawa\altaffilmark{2}}

\affil{$^1$The Institute of Statistical Mathematics,
10-3 Midori-cho, Tachikawa, Tokyo 190-8562, Japan}
\affil{$^2$Institute of Astronomy, The University of Tokyo,
2-21-1 Ohsawa, Mitaka, Tokyo 181-0015, Japan}

\begin{abstract}
Optical wide-field surveys with a high cadence
 are expected to create a new field of astronomy,  so-called ``movie astronomy,'' in the near future. 
The amount of data of the observations will  be huge, and
hence efficient data compression will be indispensable.
Here we  propose  a low-rank matrix approximation
with sparse matrix decomposition  as 
a  promising solution to reduce the data size effectively,
while preserving sufficient scientific information.
We apply one of the methods  to the movie data obtained with the prototype model of the Tomo-e Gozen
mounted on the 1.0-m Schmidt telescope of Kiso Observatory.
Once the full-scale observation of the Tomo-e Gozen commences,
it will generate $\sim$30 TB of data per night.
We demonstrate that the data are compressed by
a factor of about 10 in size without losing transient events
like optical short transient point-sources and meteors.
The intensity of point sources can be recovered
from the compressed data.
The  processing  runs sufficiently fast, compared
with the expected data-acquisition rate in the actual observing runs.
\end{abstract}




\section{Introduction} \label{sec:intro}

Optical wide-field surveys with a high cadence
has recently emerged as a potential powerful tool
to search for optical transient phenomena, 
and hence to advance  our understanding of astrophysics of compact and high-energy objects.
The Palomar Transient Factory (PTF) survey with the Palomar 1.2-m Schmidt telescope and
the Kiso Supernova Survey (KISS) with the Kiso Wide Field Camera (KWFC)
on the Kiso 1.0-m Schmidt telescope have discovered 
new species of transients with timescales of hours to days from
supernovae, novae,  to active galactic nuclei
\citep{Law+2009, Rau+2009, Sako+2012, Morokuma+2014, Tanaka+2014}.
High-cadence and wide-field monitoring surveys with the Hyper Suprime-Cam on the Subaru 8.2-m telescope
and the Kepler space telescope have successfully obtained transient light-curves
originating in shock breakouts of core-collapse supernovae
\citep{Tanaka+2016, Garnavich+2016}.
These results demonstrate importance of high-cadence wide-field survey observations
to find rare and short-duration transients.
Unsurprisingly, some next-generation  survey projects, such as 8.2-m 
Large Synoptic Survey Telescope (LSST) and Zwicky Transient Facility (ZTF)
on the Palomar Schmidt telescope,  will be also high-cadence  and cover wide fields for galactic,
extra-galactic, and solar-system objects 
\citep{Tyson_2002, Dekany+2016}.

The Tomo-e Gozen is a wide-field CMOS camera on the Kiso 1.0-m Schmidt telescope
under development at the time of writing,
and is expected to  be completed in 2018 \citep{Sako+2016, Ohsawa+2016}.
It is optimized for movie observations with sub-second to seconds time-resolutions, 
and once completed, will be capable  of taking consecutive frames with a field-of-view of
20 deg$^2$ at 2 frames per second (fps)  by 84 chips of 2k $\times$ 1k CMOS sensors.
The  primary objective of observations with the Tomo-e Gozen is to catch
rare and fast transient phenomena
with a time duration shorter than 10 seconds,
such as 
optical counterparts of fast radio bursts \citep{Keane+2016},
giant pulses in millisecond pulsars \citep{Hankins+2003},
gamma-ray bursts \citep{Gehrels_Meszaros_2012},
and binary neutron-star mergers \citep{Tanaka_Hotokezaka_2013}.
The wide-field movie observation with the Tomo-e Gozen is
also  ideal for exploring fast moving objects, 
including near-earth objects and space debris.
Such phenomena cannot be explored by the projects using CCDs
like PTF, ZTF and LSST, whose time resolutions are
60, 30 and 15 seconds, respectively.

A prototype model
of the Tomo-e Gozen with 8 CMOS chips on the Kiso Schmidt telescope
\citep[hereafter, the Tomo-e PM;][]{Sako+2016, Ohsawa+2016}
has been developed, which has achieved the frame rate of 2-fps so far.
We have obtained  movie data with a format of FITS cube for each CMOS sensor.
A data file typically consists of 400 frames of the same field images with $1136 \times 2008$ pixels.
The data rate of the Tomo-e PM is  $\sim$80 MB s$^{-1}$, corresponding to  $\sim$3 TB per night.
Then,  an observation  with the complete model of the Tomo-e Gozen
will produce a huge amount of data  of $\sim$30  TB per night.

It is therefore practically indispensable to compress the data.
The data-compression methods have two types: lossless and lossy compressions.
Since random noise pervades all the pixels in our case, the former is not effective.
For example, the popular lossless tool gzip reduced our movie data by only $\sim$18\%.
Some lossless methods were proposed for astronomical data sets \citep{Ahnen+2015, Masui+2015}.
In this paper, we discuss the latter, lossy compression.
Various lossy compression algorithms have been proposed for application
to astronomical data, such as JPEG2000 \citep{Kitaeff+2015, Vohl+2015},
the method using singular value decomposition \citep[SVD;][]{Kolev+2012}
and wavelet \citep{Belmon+2002}.
None of them, however, are so far optimized for astronomical ``movie'' data.

Commercial compression tools for movie data like MPEG4\footnote{
http://mpeg.chiariglione.org/standards/mpeg-4/video}
are readily available. However, we think that 
at least MPEG method is not appropriate for our data.
This method divides images into box regions,
and for each box region it compresses the data based
on the discrete cosine transform. The level of
compression is not uniform, 
then the fluctuation of pixel intensities becomes different
in a source and a background region used for aperture photometry.
It results in uncertainty for the photometric intensity of point sources.

Here, we propose an efficient method to compress the movie data
for both the space and time domains simultaneously in the lossy-compression manner
but without losing signals of transient phenomena,
with a calculation speed comparable or faster than the production rate
to avoid accumulating yet-unprocessed data.

\section{Method}

\begin{figure*}
 \begin{center}
   \includegraphics[width=15cm]{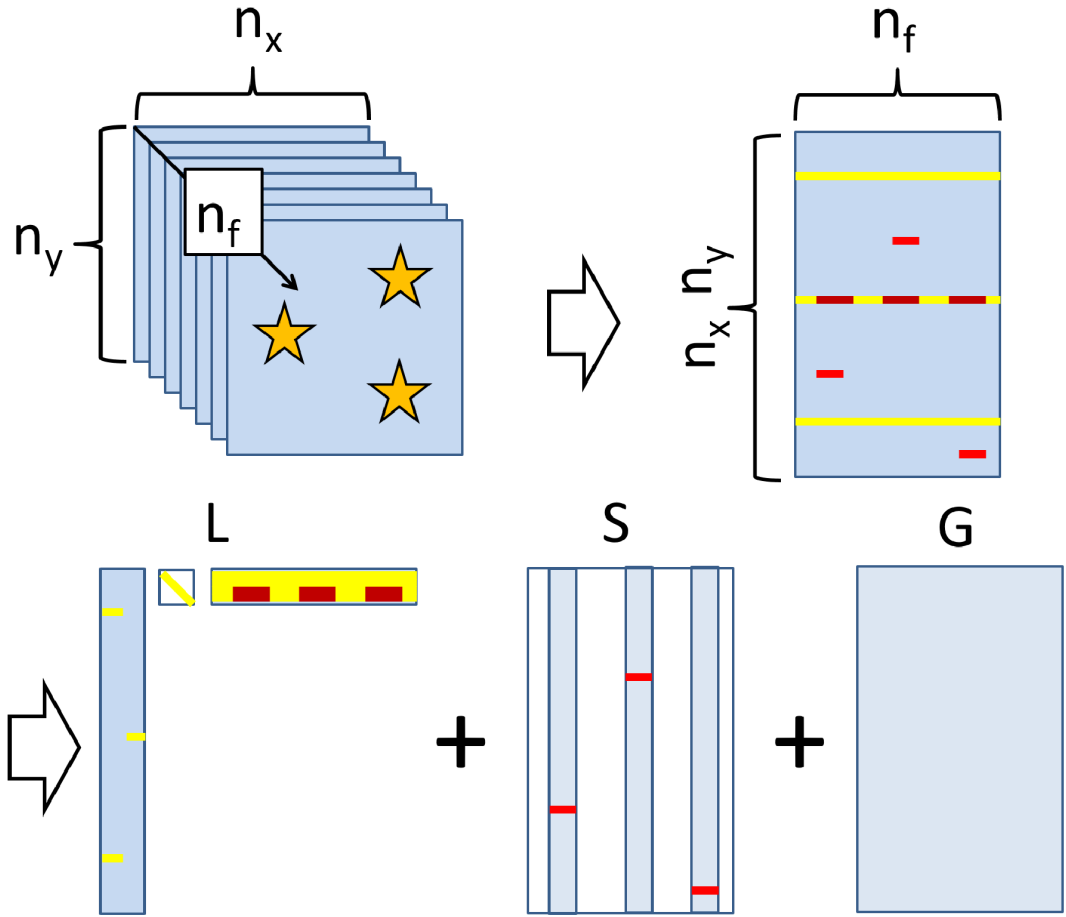}
 \end{center}
\caption{Schematic illustration of data conversion from a cube data to a matrix ($M$),
and the matrix decomposition into a low-rank ($L$), sparse ($S$) and noise matrix ($G$).
The low-rank matrix is further decomposed by SVD into $L = U D V^T$.
}\label{fig:matrix_decomp_merge}
\end{figure*}

In order to compress the movie data obtained by the Tomo-e Gozen,
we choose to use low-rank matrix approximation to reduce the data size
without losing transient events.
Before  applying it, we re-arrange the pixel values
contained in a movie data set 
($n_f$ frames of images with $n_x n_y$ pixels) into a matrix
with $n_x n_y$ rows and $n_f$ columns (Figure~\ref{fig:matrix_decomp_merge}).
Here, $n_x$, $n_y$, and $n_f$ are $1136$, $2008$, and $400$, respectively.

The concept of low-rank matrix approximation with sparse matrix
decomposition was originally proposed by \citet{Candes+2009}.
\citet{Zhou_Tao_2011} improved the algorithm and
 the computational speed, and developed GoDec.
Since the computational speed is critical in the Tomo-e Gozen application,
we develop our method based on GoDec.
In our case, an original data matrix $M$ is decomposed into $L + S + G$,
where $L$, $S$, and $G$ are a low-rank, sparse, and noise matrix, respectively
(See Figure~\ref{fig:matrix_decomp_merge}),
 so that the following function is minimized,
\begin{equation}
{\rm min}\,\, || G ||_F = {\rm min}\,\, || M - L - S ||_F
\end{equation}
subject to  ${\rm rank}(L) < r$ and ${\rm card}(S) < k$,
 where $|| \cdot ||_F$  denotes the Frobenius norm of a matrix.
The parameters $r$ and $k$ control the rank and cardinarity of
the low-rank and sparse matrices, respectively.

\begin{figure}
  \begin{center}
    \includegraphics[width=8cm]{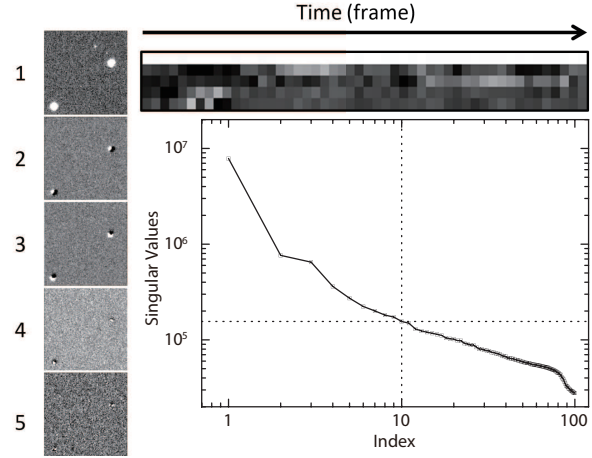}
  \end{center}
  \caption{Structure of a low-rank matrix $L$ made by the decomposition.
    The central panel shows the distribution of singular values of the matrix
    as a function of the index of the diagonal matrix $D$, obtained by SVD.
    The five images in the left-hand side show stable point sources,
    where each image is the decomposed component in the matrix $U$
    corresponding to the top five singular values.
    The upper panel shows the time variation of intensity
    of these components in the matrix $V$ in gray scale.
    \vspace{1cm}
  }\label{fig:sigval}
\end{figure}

For the low-rank matrix $L$,
we check the distribution of singular values, 
applying singular value decomposition (SVD),  as shown in the main panel of Figure~\ref{fig:sigval}.
The matrix $L$ is expressed as $L = U D V^T$, 
where $U$ and $V$ are orthogonal matrices, and $D$ is a diagonal one.
Then, we set the rank of the low-rank matrix
by setting zeros for the singular values at indices larger
than the rank.
Within the sparse matrix $S$, the transient events can be easily extracted,
because these events are innately sparse.
The time variation of the sky background can be monitored by checking the noise matrix $G$.

After the data matrix $M$ is decomposed into three matrices,
$L$, $S$, and $G$, further data processing is necessary,
because otherwise the data size would remain three times larger
than that of the original data.
The low-rank matrix $L$ is easily compressed to three small matrices,
as shown in Figure~\ref{fig:matrix_decomp_merge}.
For the sparse matrix $S$,
the frames that contain a transient event(s) must be preserved,
and the others should be discarded.
For this reduction,
we use machine-learning methods to select point sources,
which has been already established \citep[e.g.,][and references therein]{Morii+2016}.
For the noise matrix $G$, we obtain some statistics to summarize the distribution
of the pixel values and remove the matrix.

Then, the number of pixels can be reduced from
the original number of $n_{\rm org} = n_x n_y n_f$ to
\begin{equation}
n_{\rm red} = r n_x n_y + r n_f + r + n_{\rm f, sp} n_x n_y,
\end{equation}
where $r$ is a rank of the low-rank matrix $L$, and
$n_{\rm f, sp}$ is the number of frames of the sparse matrix $S$,
which contains transient  events.
Typically, $r$ is about $10$ and $n_{\rm f, sp}$ is  smaller than 10.
Therefore, the reduction factor of the data is $n_{\rm red} / n_{\rm org} \simeq 1/10$.

The original code of GoDec\footnote{The code is available at https://sites.google.com/site/godecomposition/code}
 was implemented with MatLab.
However, we  find  the original code to be impractical  to apply to our movie data
due to the speed of computation and  memory consumption.
We  have rewritten the GoDec code with C++,  utilizing
OpenBLAS\footnote{http://www.openblas.net/} and
LAPACK\footnote{http://www.netlib.org/lapack/} libraries.
We use Quick Select, instead of full sorting,
to select non-zero elements for a sparse matrix in the GoDec algorithm.

\section{Application of the proposed method}

\begin{figure*}
 \begin{center}
   \includegraphics[width=15cm]{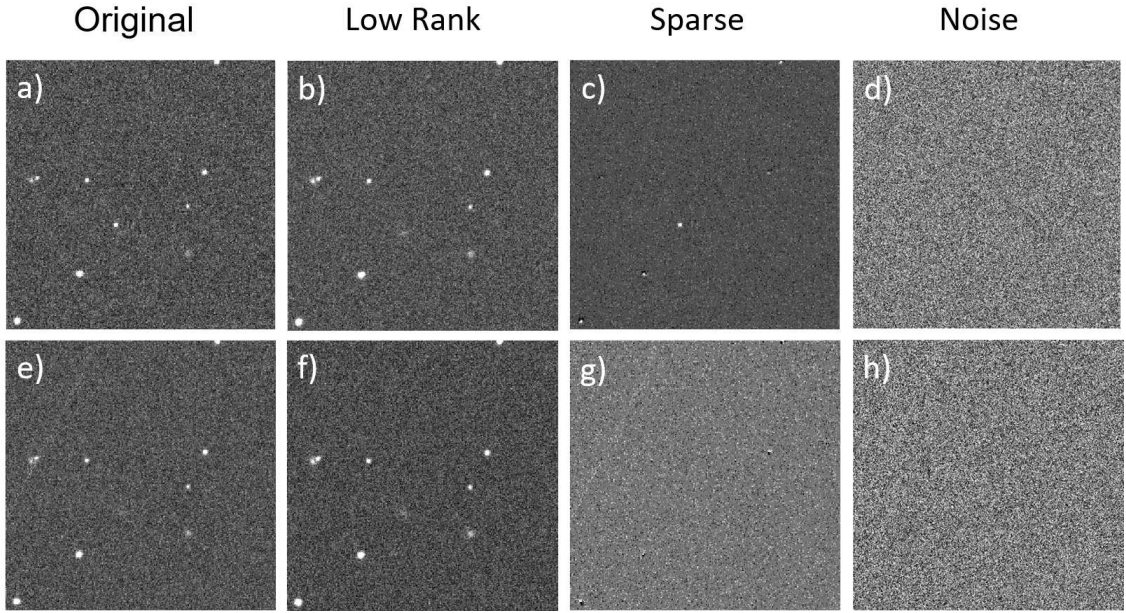}
 \end{center}
\caption{ Example decomposition images for a movie data of the Tomo-e Gozen from two frames  (top and bottom rows).
Original (denoted as the matrix $M$ in the main text), low-rank ($L$), sparse ($S$), and noise ($G$)
images are shown  in the four columns in this respective order from the left.
A transient point source  appears  near the center
of the image at the time-frame of the top row, as spotted in the original
image (a) in contrast to (e) taken in a different frame (bottom row),
and as clearly  visible in the sparse one (c) in contrast to (g).
On the other hand, a line, which is a light trail caused by a meteor,
is seen in the second time-frame (bottom row), as in the original
image (e) and sparse one (g). These transient  events are not  recognized in the low-rank
images (b, f). Noise images (d, h) do not contain any  patterns noticeable.
}\label{fig:OLS_img}
\end{figure*}

We used a movie dataset of a CMOS sensor  for 400 frames obtained 
with the Tomo-e PM  in 2015 December, which contains 
some  transient events lasting for a short duration \citep{Ohsawa+2016}.
Panels (a) and (e) of Figure~\ref{fig:OLS_img} show  sub-array images
with 300 $\times$ 300 pixels\ in two different time-frames, which 
contained a transient point source and a meteor, respectively.
We applied the decomposition to the data by setting $r = 10$
and $k = 1 \times 10^8$.
Panels (b, c, d) and (f, g, h) of Figure~\ref{fig:OLS_img}
show the result.
 A transient was extracted in the sparse matrix $S$ (panel c)
and a line  generated by a meteor was also detected (panel g).
In contrast, the low-rank image $L$ (panels b, f) did not  contain any transients.
 These results confirm that the decomposition was successful.

\begin{figure}
 \begin{center}
   \includegraphics[width=9cm]{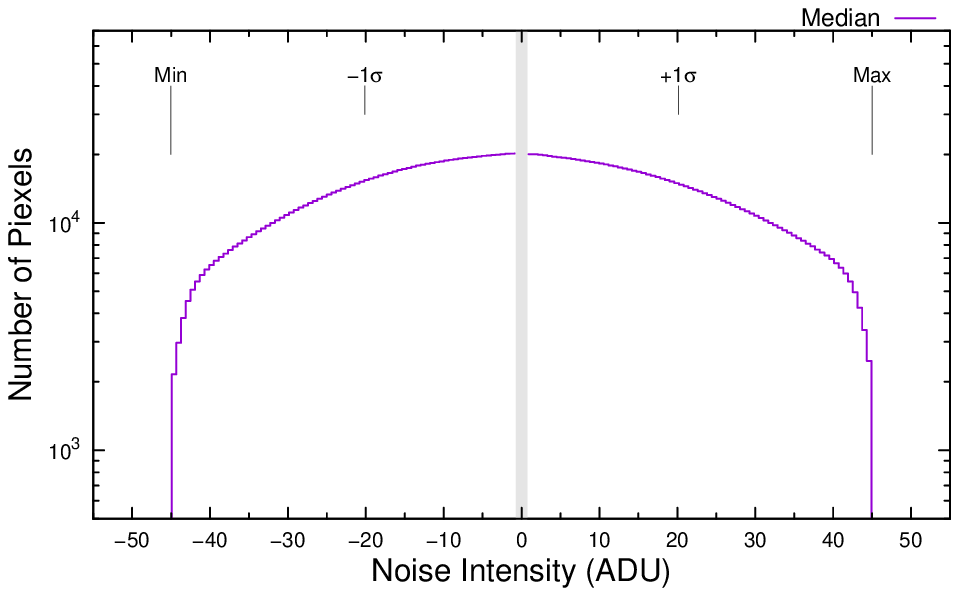}
 \end{center}
\caption{Distribution of pixel values in the noise image.
Histograms for median of pixel intensities of 400 frames are shown.
Horizontal and vertical axes are pixel intensity and number of pixels, respectively. 
The data at zero intensity is not shown.
}\label{fig:noise}
\end{figure}

\begin{figure*}
 \begin{center}
   \includegraphics[width=9cm]{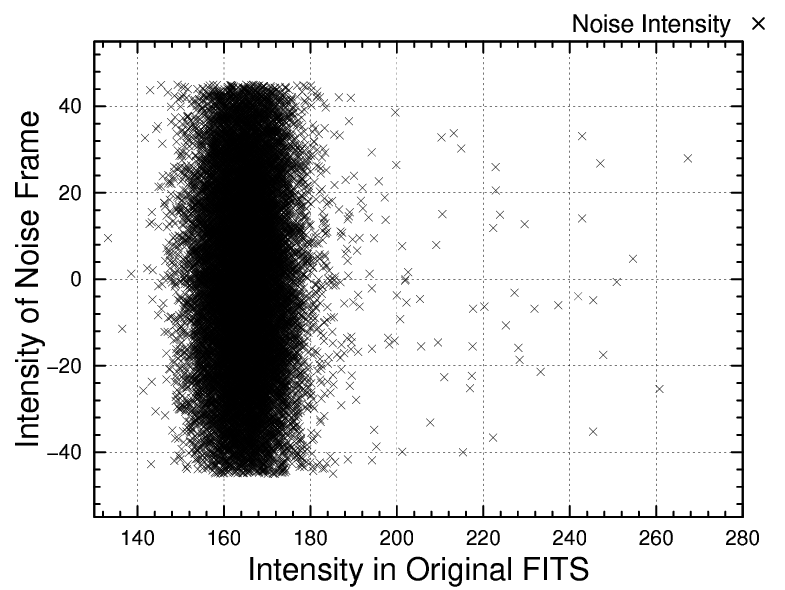}
 \end{center}
\caption{Relation between pixel intensity of noise frames $G$ and that of original frames $M$.
}\label{fig:noise_G_M}
\end{figure*}

Figure~\ref{fig:sigval} shows the singular values obtained by SVD of
the low-rank matrix $L$, suggesting that the rank of 10 is
sufficient to preserve the information of stable point sources.
Figure~\ref{fig:noise} shows the pixel values of the noise images $G$.
The histograms have a symmetrical shape, no bias and no anomaly structure.
In Figure~\ref{fig:noise_G_M}, pixel intensities of noise images $G$
were plotted against those of original images $M$,
which shows that the intensities of $G$ are independent of
the original intensities.

\begin{figure*}
 \begin{center}
   \includegraphics[width=15cm]{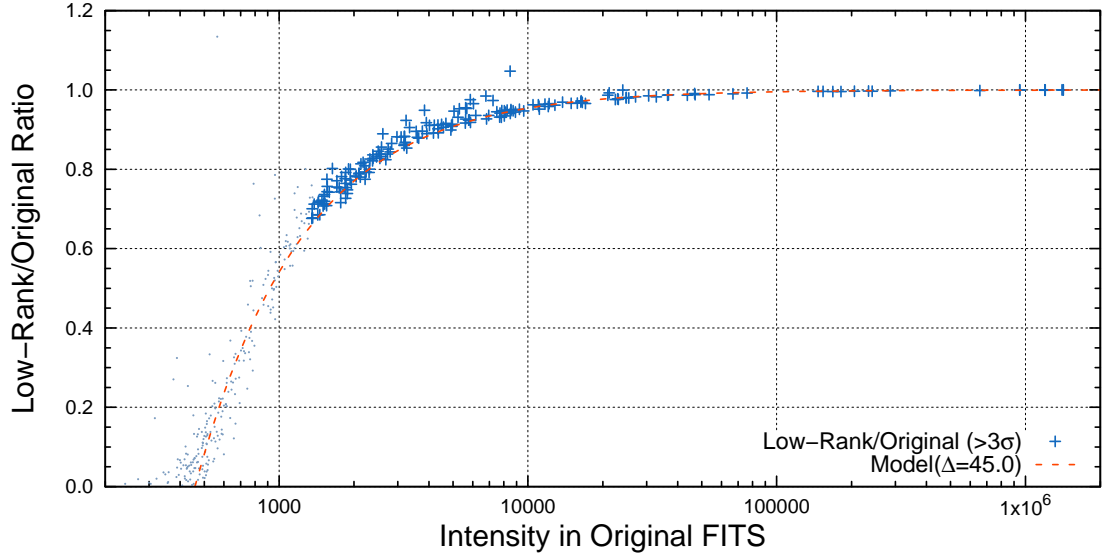}
 \end{center}
\caption{Ratio of photometric intensities of point sources in 
between the low-rank and original frames
as a function of the original intensity.
Here, the median intensity of 400 frames
obtained by photometry for each frame is shown.
Cross marks are for sources with a significance of
higher than 3 sigma level.
Red dashed-line shows the model curve for the ratio (see text).
}\label{fig:intensity_ratio}
\end{figure*}

\begin{figure}
 \begin{center}
   \includegraphics[width=9cm]{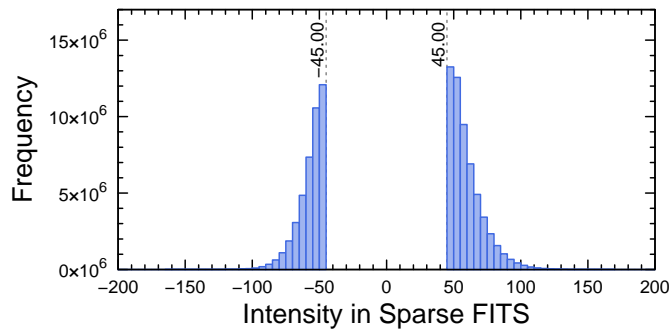}
 \end{center}
\caption{Distribution of pixel values in the sparse image.
The data at zero intensity is not shown.}\label{fig:sparse}
\end{figure}

We found that the photometric intensity of point sources in the low-rank matrix $L$
is reduced from the original value, as shown in Figure~\ref{fig:intensity_ratio}.
However, it is possible to recover the original intensity as follows.
The photometric intensity of a point source is measured by the standard aperture photometry.
To model the distribution of the pixel intensity of a point source, namely point spread function (PSF),
we assume a two-dimensional Gaussian function with symmetrical widths for simplicity.
That is, the pixel value at a distance $r$ from the center of the point source is  given by 
\begin{equation}
  {\rm PSF}(r; d; s) = s \frac{4 \ln 2}{\pi d^2} \exp\left(-4 \ln 2 \frac{r^2}{d^2} \right),
\end{equation}
where $d$ and $s$ are the FWHM and the total intensity of a point source, respectively.
Then, the  total intensity of the pixels  within an aperture radius  $R$ is
\begin{equation}
  {I_{\rm enc}}(R; d; s) = s \left[ 1 - \exp \left(-4 \ln 2 \frac{R^2}{d^2} \right) \right].
\end{equation}
For the residual from the low rank approximation,
the pixels lower than a threshold ($\Delta$) are
separated into the sparse matrix $S$ or the noise matrix $G$.
This applies for the PSF, which results in reducing the aperture size of a point source
to $r_{\Delta}$, where ${\rm PSF}(r_{\Delta}; d; s) = \Delta$.
The $\Delta$ corresponds to the lower threshold
of pixel value selected for the sparse matrix $S$.
Then, the fraction of the photometric intensity
in the low-rank matrix $L$ ($I_{\rm lr}$) to the original one
$M$ ($I_{\rm org}$) is approximated  by
\begin{equation}
\frac{I_{\rm lr}}{I_{\rm org}} \simeq \frac{{I_{\rm enc}}(r_{\Delta}; d; s)}{{I_{\rm enc}}(\infty; d; s)}
= 1 - \frac{\pi d^2 \Delta}{4 \ln 2} \frac{1}{s}.
\end{equation}
%
%
Figure~\ref{fig:intensity_ratio} shows the model curve of $I_{\rm lr} / I_{\rm org}$ for $\Delta = 45.0$,
obtained from the sparse matrix (Figure~\ref{fig:sparse}),
and it is in good agreement with the actual data.
This implies that the original intensities are recovered well from the low-rank matrix.

Next, we examined the sparse matrix, using the transient point source 
that appeared in the frame of the upper row of Figure~\ref{fig:OLS_img}.
We extracted the source from the original $M$ (panel a) and sparse $S$ (panel c) matrices,
and found that the intensity of the latter was 74\% of the former.
This would not be a problem, because
we can simply preserve the original image data
in any time-frame that is found to contain a transient source(s).

Finally, the processing time of the decomposition  was 320 s
for the movie data with $1136 \times 2008$ pixels and $400$ frames,
by setting 10 for the rank of a low-rank matrix, for which
we used a  budget computer, equipped with a CPU
of Intel Xeon Processor E5-1630 (3.70 GHz) consisting of eight cores.
The processing time is 1.6 times longer
than the duration of observation of 200 s
($0.5$ s $\times$ $400$ frames),
which is sufficiently fast for daily observations, 
as long as one CPU is used for each sensor.

\section{Discussion and Conclusion}

We have proposed to use the low-rank matrix approximation
with sparse matrix decomposition
in order to compress the movie astronomical data 
without  losing transient  events.
Compared with the conventional low-rank matrix approximation with 
principal component analysis and SVD, our method has an advantage of
preserving the transients in the sparse matrices,
which is essential for the transient-search project with the Tomo-e Gozen.

Although the value of $k$ was chosen by hand in the current study,
it should be chosen so as to maximize the number of the transients
which would be selected by the following machine-learning method. 
If the optimal $k$ is small, further data compression is possible 
for the sparse matrix $S$ by storing $2k$ numbers
for the indices and values of non-zero elements of $S$. 
In the machine-learning step for the transient detection, 
either of $S$ and $S + G$ can be used. The latter data may be
better, but we have not yet confirmed. 
We will present these studies in the future.

Our method may miss transient sources which have low significance in
each frame but are high significance by stacking
multiple frames, if they have time durations of more than a few
seconds. In our policy, we discard this type of sources,
because the primary objective of the Tomo-e Gozen project is
discovering transient sources with a short time scale. In addition, it
is possible to keep these transients by stacking raw frames in various
time durations like 5, 10, 50 seconds and so on, 
before applying our compression method.

Now, we are  developing both the hardware and software systems of the Tomo-e Gozen survey,
including the CMOS camera, data analysis pipeline, and scheduling.
Our proposed method  has successfully overcome one of the  challenges  that the project faces.
The Tomo-e Gozen will pioneer  a new field of astronomy, ``movie astronomy.'' 

\acknowledgments

This research is supported by CREST
and PRESTO, Japan Science and Technology Agency (JST), and
 in part, by JSPS Grants-in-Aid for Scientific Research (KAKENHI)
Grant Number JP25120008, JP25103502, JP26247074, JP24103001, JP16H02158 and JP16H06341.
This work was achieved using the grant of Joint Development Research
by the Research Coordination Committee, National Astronomical Observatory of Japan (NAOJ).
The  Tomo-e PM was  developed in collaboration
with the Advanced Technology Center of NAOJ and
 is equipped with the full-HD CMOS sensors developed in collaboration with Canon Inc.

\end{document}